\documentclass[conference]{IEEEtran}
\IEEEoverridecommandlockouts
\usepackage{cite}
\usepackage{amsmath,amssymb,amsfonts}
\usepackage{algorithmic}
\usepackage{graphicx}
\usepackage{textcomp}
\usepackage{xcolor}
\usepackage{siunitx}
\usepackage{tabularx} 
\usepackage{booktabs} 
\usepackage{ragged2e} 
\usepackage{makecell}
\newcommand{\spellbreak}[2]{\texttt{#1}\newline\texttt{#2}}

\def\BibTeX{{\rm B\kern-.05em{\sc i\kern-.025em b}\kern-.08em
    T\kern-.1667em\lower.7ex\hbox{E}\kern-.125emX}}
\begin{document}

\title{Decoding Workload and
Agreement From EEG During Spoken Dialogue With Conversational AI\\
}

\author{
\IEEEauthorblockN{Lucija Mihić Zidar}
\IEEEauthorblockA{
\textit{Chair of Neuroadaptive Human--Computer Interaction} \\
\textit{Brandenburg Technical University Cottbus--Senftenberg} \\
\textit{Auryal GmbH} \\
Cottbus, Germany \\
lucija.mihiczidar@b-tu.de
}
\and
\IEEEauthorblockN{Philipp Wicke}
\IEEEauthorblockA{
\textit{Auryal GmbH} \\
Berlin, Germany \\
philipp@auryal.com
}
\and
\IEEEauthorblockN{Praneel Bhatia}
\IEEEauthorblockA{
\textit{Auryal GmbH} \\
Berlin, Germany \\
praneel@auryal.com
}
\and
\IEEEauthorblockN{Rosa Lutz}
\IEEEauthorblockA{
\textit{Auryal GmbH} \\
Berlin, Germany \\
rosa@auryal.com
}
\and
\IEEEauthorblockN{Marius Klug}
\IEEEauthorblockA{
\textit{Chair of Neuroadaptive Human--Computer Interaction} \\
\textit{Brandenburg Technical University Cottbus--Senftenberg} \\
\textit{Auryal GmbH} \\
Cottbus, Germany \\
marius.klug@b-tu.de
}
\and
\IEEEauthorblockN{Thorsten O. Zander}
\IEEEauthorblockA{
\textit{Chair of Neuroadaptive Human--Computer Interaction} \\
\textit{Brandenburg Technical University Cottbus--Senftenberg} \\
Cottbus, Germany \\
thorsten.zander@b-tu.de
}
}


\maketitle

\begin{abstract}
Passive brain–computer interfaces offer a potential source of implicit feedback for alignment of large language models, but most mental state decoding has been done in controlled tasks. This paper investigates whether established EEG classifiers for mental workload and implicit agreement can be transferred to spoken human–AI dialogue. We introduce two conversational paradigms – a Spelling Bee task and a sentence-completion task – and an end-to-end pipeline for transcribing, annotating, and aligning word-level conversational events with continuous EEG classifier output. In a pilot study, workload decoding showed interpretable trends during spoken interaction, supporting cross-paradigm transfer. For implicit agreement, we demonstrate continuous application and precise temporal alignment to conversational events, while identifying limitations related to construct transfer and asynchronous application of event-based classifiers. Overall, the results establish feasibility and constraints for integrating passive BCI signals into conversational AI systems.
\end{abstract}


\section{Introduction}

Conversational AI, built around Large Language Models (LLMs), has rapidly become ubiquitous in everyday life, increasingly mediating how people search for information, perform work tasks, seek personal advice, and interact with digital systems in general \cite{ling2021factors}. Alongside text-based interfaces, many contemporary systems now support spoken dialogue through integrated, real-time speech recognition and synthesis, enabling more natural and human-like conversations \cite{van2016wavenet,dinh2025benchmarking}. As these systems become more pervasive and socially embedded, ensuring their behavior aligns with human expectations and intentions becomes increasingly critical.

At present, LLM alignment - that is, the process of shaping models to behave in accordance with human values, intentions, and preferences - relies predominantly on reinforcement learning from explicit human feedback (RLHF), where users or annotators provide preference judgments, rankings, or evaluations of model outputs \cite{Kaufmann2023-ru, Ziegler2019-rj}. While effective for shaping coarse aspects of model behavior, this form of explicit feedback is inherently low-bandwidth and relies on retrospective human judgments. Therefore, it remains limited in its ability to reflect users’ momentarily cognitive and affective experiences during the interaction. As a result, a full range of subtle and nuanced reactions remain largely invisible to the system, limiting its capacity for fine-grained, real-time adaptation \cite{casper2023open}.

Implicit feedback, derived from naturally occurring behavioral, physiological or cognitive responses during interaction, offers a complementary way to access this information without requiring additional human effort \cite{Kaufmann2025-ds}. One such approach explored in recent work is the use of electroencephalography (EEG)-based passive brain-computer interfaces (pBCIs) as a source of implicit feedback for LLM alignment \cite{Gherman2025-dl}. By decoding spontaneous neural activity that arises during ongoing interaction, pBCIs enable inference of users’ internal states without requiring any intentional action on the part of the user \cite{Zander2011-ez}. Our work addresses two mental states particularly relevant for human-LLM interaction: mental workload and implicit evaluation or ``agreement'' to the system's action. 

Mental workload reflects the cognitive demands placed on the user and is a central construct in human factors and neuroergonomics \cite{Wickens2008-fl}. In EEG, workload is typically associated with increased frontal theta and decreased parietal alpha power \cite{Klimesch1999-wh}. Workload BCIs have been applied to various task types, including n-back, dual-task paradigms, attention and memory tasks \cite{Hassan2025-um}, as well complex real-world scenarios, such as aviation \cite{hamann2022investigating} and surgery \cite{Zander2017-rl}.
Importantly, workload classifiers have been found to generalize across domains. A model trained on arithmetic data has been successfully applied across diverse task contexts, including mental rotation, span tasks \cite{Krol2016-my}, and text reading \cite{Andreessen2021-oq}.

The second mental state addressed in this paper is the user’s ``agreement'' to system's behavior, reflecting whether an observed action aligns with perceived expectations in a given context. This construct has been introduced in \cite{Zander2016-rb} using a paradigm, in which participants observed a cursor jumps on a grid with a predefined target. Their implicit evaluations of cursor movements with respect to goal-congruency were decoded from single-trial event-related potentials (ERPs) originating predominantly in the medial prefrontal cortex, with amplitudes linearly corresponding to the degree of expectation violation. 

Despite substantial progress in decoding mental states from EEG, most pBCI classifiers have been developed and validated in controlled paradigms with clearly defined tasks and precisely time-locked events. Extending these approaches to more realistic, interactive scenarios, such as spoken dialogue with an AI agent, poses significant challenges, particularly for event-related mental states that rely on well-defined stimulus onsets and reliable ground-truth labeling \cite{Spuler2015-xp}. 


This paper examines whether established pBCI classifiers for mental workload and agreement, originally developed in controlled tasks, can be applied to spoken human–AI interaction - so that these signals can be used for additional feedback or training of AI. To this end, we designed two conversational paradigms to elicit mental workload and agreement in a spoken interaction with an AI agent and implemented a pipeline that aligns continuous EEG-based classifier output with word-level conversational events. Based on two pilot studies, we present an initial feasibility assessment of applying workload and agreement classifiers in conversational settings and identify boundary conditions for their transfer to naturalistic dialogue.

\section{Methods}

\subsection{Participants and procedure}
Four male participants with a mean age of 30.3 years ($SD=7.3$) took part in our study. Prior to the experiment, all participants gave their written informed consent, and were compensated with 15 EUR per hour of participation. Each participant completed one session, which involved a calibration paradigm, followed by a conversational paradigm, both related to one mental state. Two participants were recorded for workload paradigms and two participants were recorded for agreement paradigms. Experiment duration excluding preparation and self-paced breaks was $\sim$25 min for the workload paradigms (8 min calibration, 15 min conversational task) and $\sim$45 min for the agreement paradigms (9 min calibration, 35 min conversational task).

\subsection{Calibration paradigms}

For the calibration of the workload predictive model we used a paradigm introduced in \cite{Krol2016-my}, in which participants alternated between 10-second phases of high and low workload. In the high-workload phase, a subtraction problem was shown on the screen, and participants continuously subtracted the smaller number from the larger. In the low-workload phase, a fixation cross was shown and participants were instructed to relax and think of a pleasant memory while keeping their eyes open. In both phases, slowly moving glowing spots were presented on the screen in 50\% of the trials to equalize ocular artifacts between classes. Participants completed 20 trials per condition with a self-paced break halfway through.

To calibrate the agreement predictive model, we adapted the grid-navigation paradigm introduced in \cite{Zander2016-rb}. Participants viewed a 4x4 gray grid on a black background with a designated target node in one corner and a red cursor positioned at one of the other nodes. On each trial, the cursor performed a single jump to a random adjacent node: a 1 second countdown animation preceded the movement, followed by a 1 sec display of the new position with the traversed grid line highlighted, and an additional 1 sec pause before the next trial. Participants were instructed to mentally evaluate each cursor movement as ``acceptable'' or ``not acceptable'' with respect to reaching the target. Participants completed three blocks of 178 trials each with self-paced breaks in between.

\subsection{Conversational paradigms}
Two conversational paradigms were developed to elicit neural correlates of agreement and workload during spoken human–AI interaction. In both paradigms, participants interacted verbally with a conversational AI agent (the former with \textit{OpenAI} Realtime\footnote{https://platform.openai.com/docs/guides/realtime} and the latter with \textit{Google} Gemini 2.0 Flash\footnote{https://gemini.google.com/} and speech synthesis via the platform ElevenLabs\footnote{https://elevenlabs.io/}).

\subsubsection{Workload Paradigm}

Participants completed an AI-guided, 10-round Spelling Bee moderated by conversational AI agent. In each round, the agent presented one target word and required (i) a pronunciation attempt and (ii) a spelling attempt; words increased in difficulty across rounds to elicit a broad range of task demands. The list of words was fixed for all participants. Participant speech was transcribed online using a speed-optimised, Whisper-based \cite{radford2023robust} ASR\footnote{https://github.com/SYSTRAN/faster-whisper} and judged in real time by a realtime LLM voice agent (OpenAI Realtime API; gpt-realtime) that returned structured feedback (Correct/Wrong) plus a brief explanation and advanced immediately to the next round. To reduce interruptions during ongoing spelling (consistent with conversational turn-taking constraints), we increased server-side silence thresholds. The 10-word list was generated, once a priori via iterative prompt refinement on GPT-4o, and two rounds implemented pre-specified, intentional misjudgments (Rounds 4 and 7) to introduce standardized agent errors to be able to apply an error BCI classifier in a follow up investigation. Audio I/O and conversation logging were handled by the web interface, and transcripts were extracted from the session logs post hoc; to avoid instability from residual agent state, a fresh agent instance was created for each participant session. Table~\ref{tab:spellingbee_examples} illustrates representative rounds, showing the required pronunciation/spelling attempts and the agent’s fixed feedback format.

\begin{table}[htbp]
\caption{Example Spelling Bee interaction rounds.}
\centering
\scriptsize
\setlength{\tabcolsep}{3pt}
\renewcommand{\arraystretch}{1.15}
\label{tab:spellingbee_examples}
\begin{tabularx}{\linewidth}{@{} c l >{\RaggedRight\arraybackslash}X >{\RaggedRight\arraybackslash}X @{}}
\toprule
\textbf{Rnd} & \textbf{Word} & \textbf{Participant attempts} & \textbf{Agent feedback} \\ \midrule
1 & Book &
P: \textit{``book''}\newline
S: \texttt{B-O-O-K} &
\textcolor{white}{Filler} \,$\rightarrow$\,\texttt{...}\,$\rightarrow$\,\textbf{Correct!}\newline
\textit{``...because the word is spelled BOOK.''} \\ \addlinespace

5 & Rendezvous &
P: \textit{``RON-day-voo''}\newline
S: \spellbreak{R-E-N-D-E-Z-V-O}{-U-S} &
\textcolor{white}{Filler} \,$\rightarrow$\,\texttt{...}\,$\rightarrow$\,\textbf{Correct!}\newline
\textit{``...because the word is spelled RENDEZVOUS.''} \\ \addlinespace

9 & \makecell[l]{Antidisestablish-\\mentarianism} &
P: \textit{``an-tee-dis-uh-STAB-lish-men-TARE-ee-uh-niz-um''}\newline
S: \texttt{A-N-T-I-D-I-S-E}\newline
\texttt{-S-T-A-B-L-I-S-H}\newline
\texttt{-M-E-N-T-A-R-I}\newline
\texttt{-A-N-I-S-M} &
\textcolor{white}{Filler} \,$\rightarrow$\,\texttt{...}\,$\rightarrow$\,\textbf{Wrong}\newline
\textit{``...because the correct spelling includes \texttt{...arianism} at the end.''} \\ \bottomrule
\end{tabularx}
\end{table}


\subsubsection{Agreement Paradigm}

The agreement calibration is measuring the response to agreement and disagreement in a spatial setting (jump on a grid away from or towards an expected target). We leverage distributional semantics and semantic vectorization of words in a sentence completion task to induce an analogous response in the conversational paradigm. 

\begin{figure}
    \centering
    \includegraphics[width=0.7\linewidth]{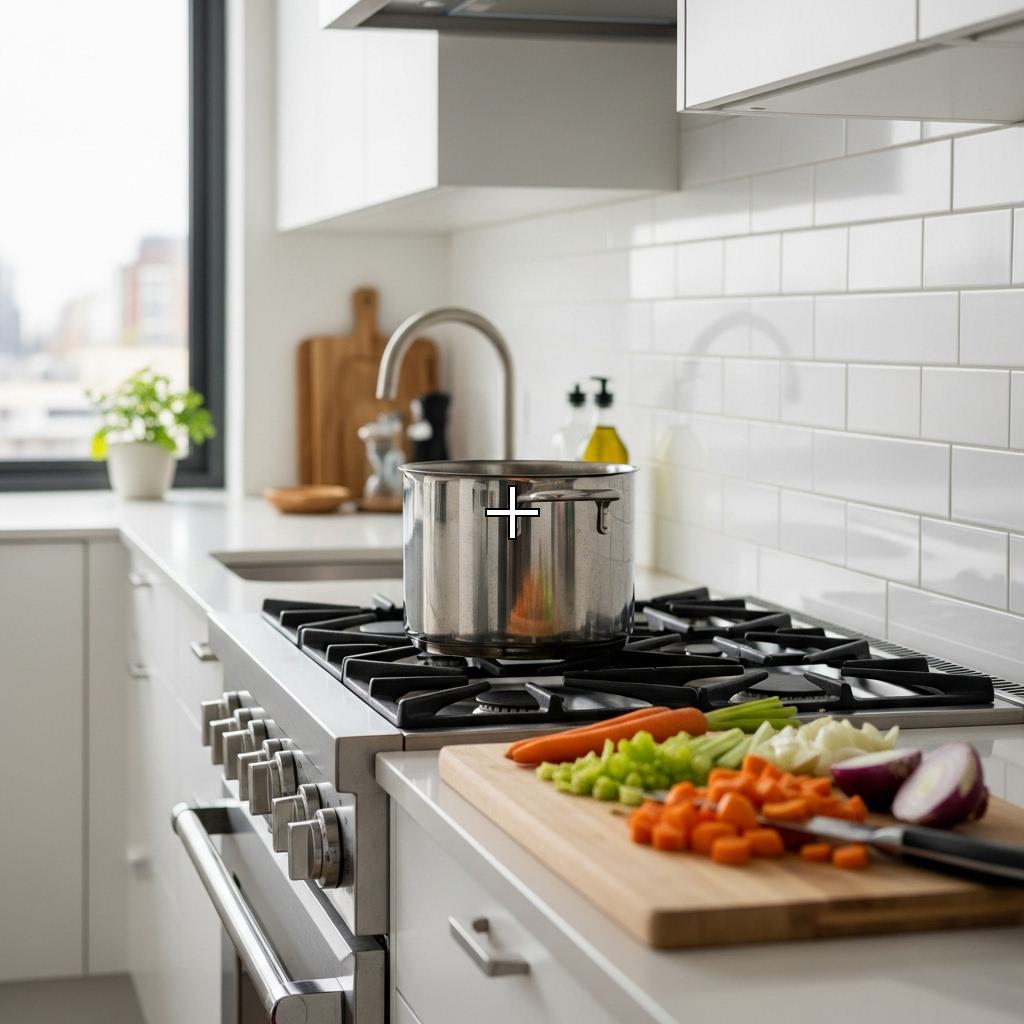}
    \caption{Example scene stimulus (kitchen) used in the agreement paradigm.}

    \label{fig:agreement_img04}
\end{figure}

Participants completed an AI-guided dialogue task in which a conversational agent sequentially presented 10 photorealistic, AI-generated scene images (each overlaid with a central fixation cross to stabilize gaze during concurrent EEG recording; \cite{dimigen2011coregistration}) and, per scene, six sentence-completion trials (60 total). An example scene stimulus is shown in Figure~\ref{fig:agreement_img04}. Images were created using \textit{Google} \texttt{imagen-4.0} and iterative prompt refinement. Sentence stems were sampled from large-scale English cloze norms \cite{peelle2020completion} and paired with a single displayed completion word intended to elicit high, medium, or low contextual agreement. High-agreement completions were the modal cloze responses from the norms \cite{peelle2020completion}; medium- and low-agreement completions were constructed to be progressively less context-appropriate using distributional-semantic similarity (cosine distance) based on pretrained word/sentence embedding methods (word2vec; Sentence-BERT/MPNet) \cite{mikolov2013efficient,reimers2019sentencebert}. On each trial, participants provided verbal responses, including an unexpectedness rating (1--5 Likert scale) and a report of the expected word. Table~\ref{tab:dialogue_examples} shows representative high/medium/low agreement trials and response fields. Candidate items were LLM-filtered from the 3{,}085 normed contexts \cite{peelle2020completion}, automatically cross-checked against the source dataset to prevent selection inconsistencies, and residual sentence-level errors were corrected manually.

\begin{table}[htbp]
\caption{Example verbal interaction trials.}
\centering
\small 
\label{tab:dialogue_examples}
\begin{tabularx}{\linewidth}{@{} l >{\RaggedRight}X c l @{}}
\toprule
\textbf{Agreement} & \textbf{AI Stimulus Sentence} & \textbf{Rating} & \textbf{Expected Word} \\ \midrule
High & She chopped onions before adding them to the \textbf{soup}. & 5 & \textit{expected} (soup) \\ \addlinespace
Medium & Before cooking, the woman had to clean the \textbf{grease}. & 3 & stove \\ \addlinespace
Low & While eating the soup, Ariel could taste the \textbf{mold}. & 1 & salt \\ \bottomrule
\multicolumn{4}{@{}p{\linewidth}@{}}{\footnotesize \centering{Rating scale: 1 = ``completely unexpected'', 5 = ``completely expected''.}}
\end{tabularx}
\end{table}

\subsection{Data acquisition}
EEG was recorded with 64 active actiCAP slim gel electrodes (Brain Products GmbH, Gilching, Germany) positioned according to the extended 10-20 international system \cite{Klem1999-td}. Data were sampled at 500 Hz and amplified with an ActiCHamp amplifier (Brain Products GmbH, Gilching, Germany). EEG recording was done reference-free with Fpz as the ground electrode. During gelling, the impedances were kept under 20 k$\Omega$. Earphones and a PC microphone were used for the voice chat interaction. During the conversational paradigm, OBS Studio\footnote{https://obsproject.com/} recorded the computer audio, the participants’ speech via a PC 
microphone, a webcam view of the participant, and the on-screen content.
All data streams were synchronized and recorded via the Lab Streaming Layer (LSL) \cite{Kothe2024-yu}. 

\subsection{Transcription and annotation}
In addition to EEG, we pre-processed audio/video and transcripts to enable time-resolved annotation and multimodal alignment. Speech was transcribed and force-aligned at the word level using \texttt{faster\_whisper}\footnote{https://github.com/SYSTRAN/faster-whisper}, yielding word-level timestamps that were used to estimate and correct temporal offsets between the EEG and audio streams and to export synchronized timing metadata for downstream analysis. The aligned transcripts were then used to generate captioned experiment videos utilising OCR with an on-screen timer - and, for visualization/quality control, to render classifier-overlay videos. For semantic labeling, two annotators (co-authors) independently segmented the interaction into the 10 task rounds; based on this agreement, subsequent analyses were conducted on round-based segments to characterize workload and agreement changes across the session.

\subsection{EEG processing}

EEG data were processed in MATLAB R2024a using EEGLAB v2024.2.1 \cite{Delorme2004-ww}. EEG was first downsampled to 250 Hz and re-referenced to the average reference. For each participant, datasets from the calibration and conversational paradigms were merged to enable joint preprocessing. To prepare the data for independent component analysis (ICA), we applied a 1 Hz high-pass filter and performed automated time-domain artifact rejection using the BeMoBIL pipeline \cite{Klug2022-bb}, which removed the worst 2\% of time windows based on a multivariate combination of amplitude, variance, and spectral criteria. The noise-cleaned data were then decomposed using the AMICA algorithm \cite{Palmer2011-og}.
Independent components were automatically classified using ICLabel \cite{Pion-Tonachini2019-ea} (lite classifier). Components whose highest-probability label corresponded to non-brain sources were flagged for rejection, retaining only components labeled as brain or other. The resulting ICA weights, sphere, and ICLabel classifications were transferred back to the original (non-merged) re-referenced datasets, where non-brain components were removed before further analysis.

\subsection{Calibration of workload and agreement predictive models}

All analyses were based on subject-specific BCI models trained offline using BCILAB 1.4-devel \cite{Kothe2013-et}. Separate classifiers were calibrated for mental workload and implicit agreement following approaches from their respective original studies (for more detail see \cite{Krol2016-my} for workload and \cite{Zander2016-rb} for agreement).

EEG data recorded during the arithmetic task were used to train a workload classifier using filter bank common spatial patterns (fbCSP) \cite{Ang2008-sb}. Spectral features were extracted from the theta (4–7 Hz) and alpha (8–13 Hz) bands. The EEG was resampled to 100 Hz, segmented into 1-s epochs, and projected through the learned spatial filters. These features served as input to a regularized linear discriminant analysis (LDA) classifier to discriminate between low and high workload trials. 

Agreement classifier was calibrated on EEG data from the grid task. Trials were labeled based on the angular deviance between the cursor's path and the straight line to the target: $< 45^{\circ}$ for  ``correct'' and $> 90^{\circ}$ for ``incorrect'' jumps. This resulted in approximately balanced classes, with a mean of 85.5 ($SD=2.9$) trials per class.
For feature extraction, EEG data were resampled to 100 Hz and band-pass filtered between 0.1 and 15 Hz. Event-related features were computed using a windowed-means approach \cite{Blankertz2011-jv} from non-overlapping 50 ms windows spanning 200–650 ms after each jump. Regularized LDA was used to discriminate between correct and incorrect jumps.

Calibration accuracy of the workload and agreement predictive models was estimated for each participant using repeated 5×5 cross-validation procedure. Chance accuracy was determined by simulating a random classifier that guessed class labels in proportion to the observed class frequencies. The upper bound of the one-sided 95\% Wilson confidence interval was used as the significance threshold \cite{billinger2012significant}.

\subsection{Workload and agreement decoding in conversational paradigms}

For continuous evaluation of mental workload and implicit agreement in conversational paradigms, we applied the individually calibrated classifiers to corresponding conversational paradigms by simulating online classification at a frequency of 50 Hz with BCILAB’s \texttt{onl\_simulate} function. The outputs of both models were continuous prediction values, with a decision boundary centered at zero. These prediction values were subsequently time-aligned with audio-derived word onset and offset timestamps.

Further quantitative analysis of classifier output applied to conversational data was limited to the workload classifier. To assess changes in workload across Spelling Bee rounds, predicted workload values were aggregated within each round to obtain a mean workload value per round for each participant. To account for temporal autocorrelation within rounds, uncertainty of the round-level means was estimated using an AR(1)-adjusted procedure. Statistical significance was assessed using two-sided 95\% confidence intervals of the ordinary least squares (OLS) slope, with significance defined as confidence intervals excluding zero (\(\alpha = 0.05\)).

To verify continuous application of the event-based agreement classifier, we first applied it to its own calibration data to confirm expected behavior under controlled, time-locked conditions. The classifier output exhibited continuous fluctuations rather than discrete, event-locked responses. Given this behavior, further statistical evaluation of agreement decoding in the conversational paradigm was not pursued in the present study.

\section{Results}

For workload calibration, the classification accuracies of both participants significantly exceeded the chance threshold of 54.1\%, with a mean accuracy of 67.8\% ($SD=5.3$) for participant 1 and  81.0\% ($SD=1.1\%$) for participant 2.

The mean agreement cross-validated accuracies were also significantly above the chance threshold of 56.2\%, with a mean accuracy of 67.4\% ($SD=8.9\%$) for participant 3 and 64.1\% ($SD=9.4\%$) for participant 4.

\begin{figure}
    \centering
    \includegraphics[width=1\linewidth]{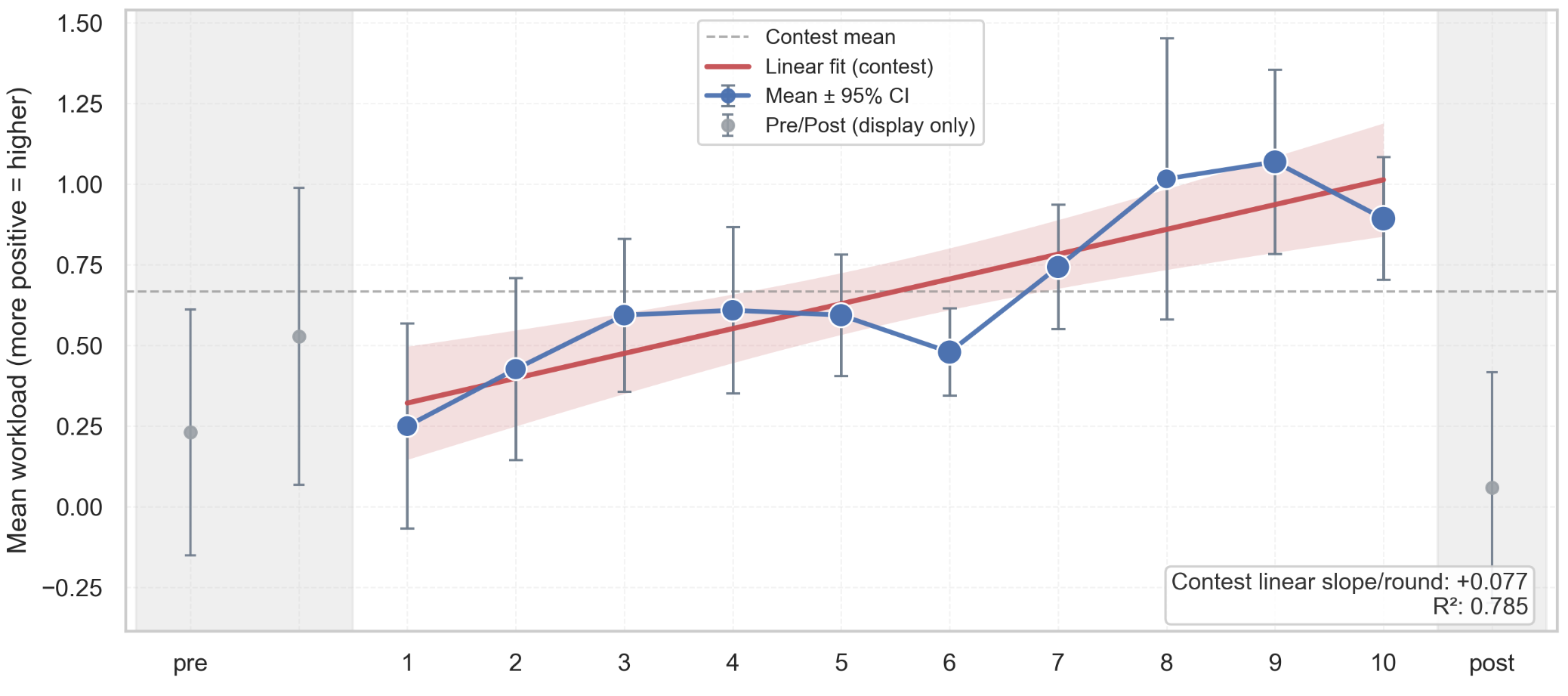}
    \caption{Participant \#1 - Round-level Workload in the Spelling Bee Paradigm (Means with AR(1)-aware 95\% CI)}
    \label{fig:part001WL}
\end{figure}

\begin{figure}
    \centering
    \includegraphics[width=1\linewidth]{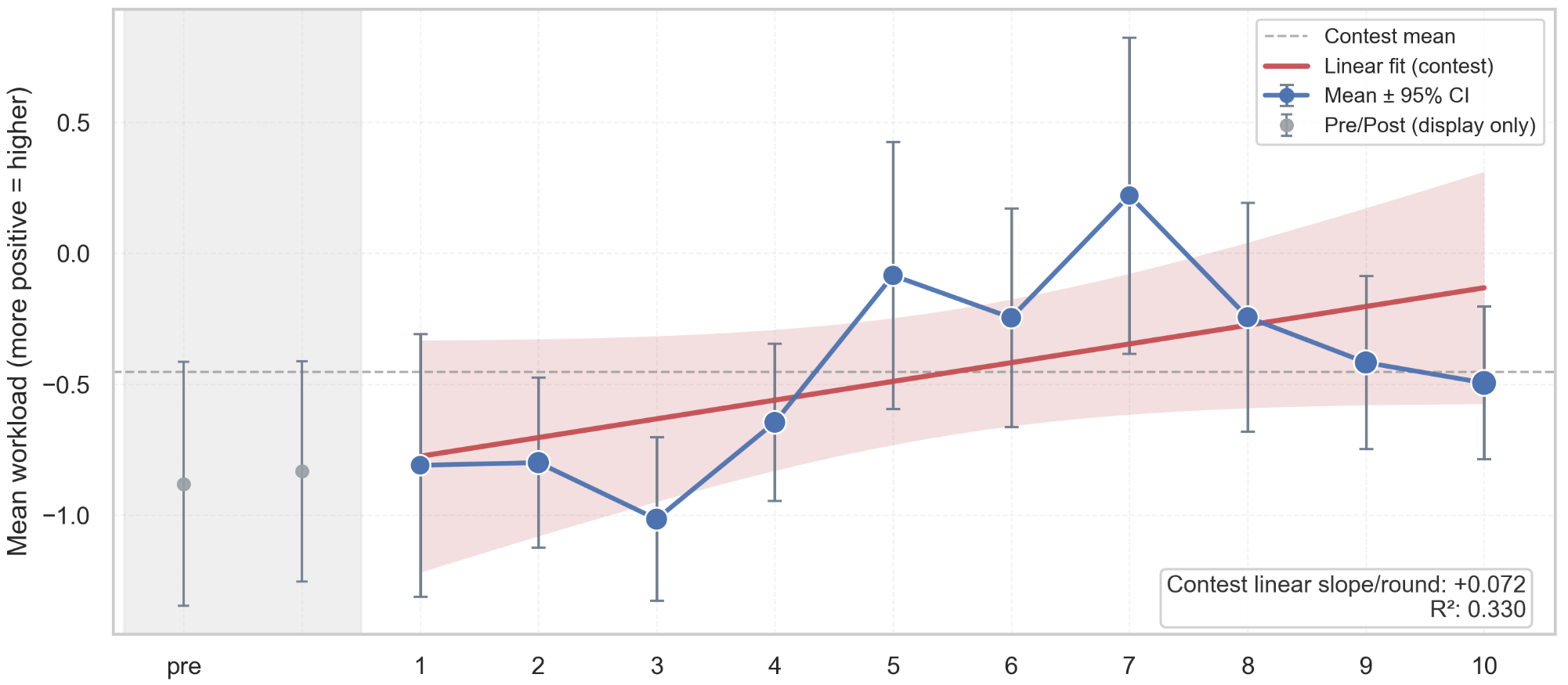}
    \caption{Participant \#2 - Round-level Workload in the Spelling Bee Paradigm (Means with AR(1)-aware 95\% CI)}
    \label{fig:part002WL}
\end{figure}

Across two pilot sessions, decoded workload tended to increase with spelling-bee round number, consistent with increasing task difficulty. 
For participant 1, workload increased consistently across rounds (see fig.~\ref{fig:part001WL}). OLS regression revealed a statistically significant positive linear trend, with an estimated slope of $+0.08$ (95\% CI [$0.04$, $0.11$], $p < 0.001$, $R^2 = 0.79$). This indicates that approximately 79\% of the between-round variance in mean workload was explained by a linear increase over rounds. The cumulative change implied by the model corresponds to an increase of predicted workload values by $+0.69$ from round 1 to round 10.
For participant~2, the estimated linear trend was positive but did not reach statistical significance. As shown in Fig.~\ref{fig:part002WL}, round-level workload increased over the early and middle rounds, peaking around round~7, followed by a decrease in later rounds. The OLS slope was $+0.07$ per round (95\% CI [$-0.01$, $0.16$], $p = 0.08$, $R^2 = 0.33$), suggesting an overall upward, but statistically non-significant trend. The cumulative change of predicted workload from round~1 to round~10 was $+0.315$ (95\% CI [$-0.266$, $0.895$]), consistent with a positive but statistically inconclusive overall increase.

\begin{figure}
    \centering
    \includegraphics[width=1\linewidth]{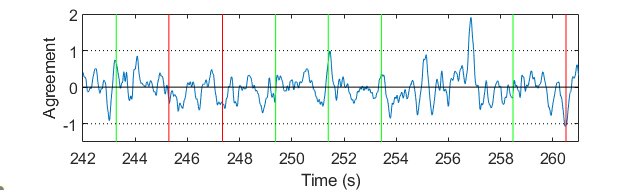}
    \caption{Participant \#3 - Agreement classifier output applied to its own training data from the grid task (raw predictive values smoothed with a moving average over 10 samples, decision boundary at 0). Correct and incorrect jump events marked with green and red, respectively.}
    \label{fig:part003_AG}
\end{figure}

\begin{figure}
    \centering
    \includegraphics[width=1\linewidth]{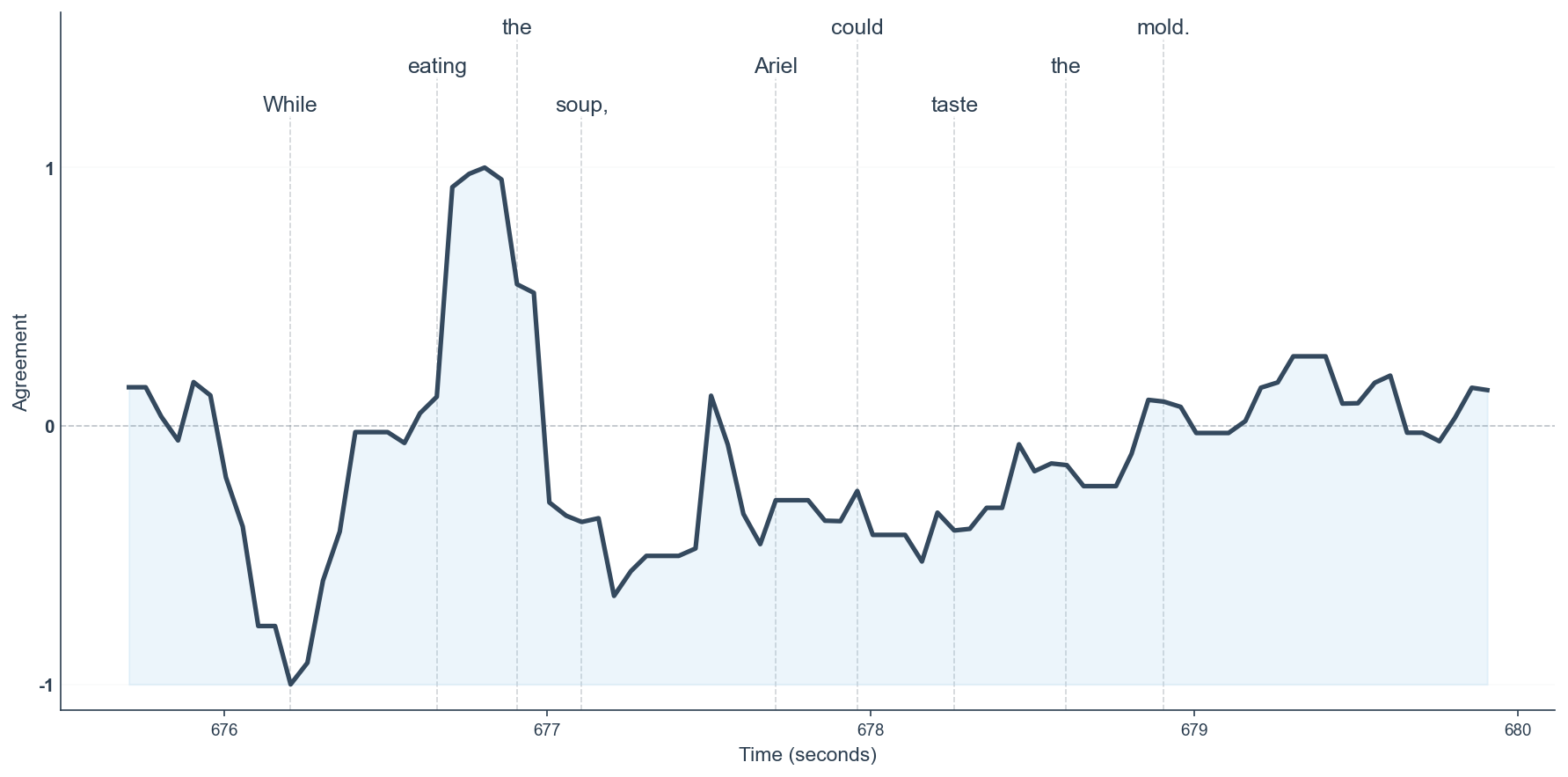}
    \caption{Participant \#3 - Agreement classifier output applied to the conversational paradigm (values normalized from -1 to 1, decision boundary at 0) with timestamps of word onset.}
    \label{fig:part003_AG_conv}
\end{figure}

For agreement, we show preliminary results illustrating the behavior of the classifier under continuous application.
Figure~\ref{fig:part003_AG} shows the agreement classifier applied continuously to the grid-task data, with cursor-movement events marked. Figure~\ref{fig:part003_AG_conv} shows the same classifier applied to the conversational paradigm, time-aligned to word onset. In both cases, the classifier output exhibits continuous fluctuations rather than remaining near the decision boundary outside experimentally defined events.

\section{Discussion}

This study examined whether EEG-based passive BCI classifiers for mental workload and implicit evaluation can be transferred from controlled laboratory paradigms to a naturalistic, spoken human-AI interaction. The main contribution of this work is a feasibility demonstration: we introduce two spoken-dialogue paradigms aiming to bridge controlled and naturalistic paradigms, and we implement a pipeline that enables transcription, annotation, and alignment of word-level timestamps with continuous EEG-based classifier output.

In the Spelling Bee paradigm, decoded workload tended to increase with task progression. For one participant this rise was statistically significant across rounds, indicating that a workload classifier trained on an arithmetic calibration task can generalize to spoken interaction with an AI agent. For the second participant the trend was positive but non-significant and decreased in the final rounds, plausibly reflecting disengagement under excessive difficulty. 

We further examined whether an implicit agreement classifier, calibrated in a spatial grid-navigation paradigm, transfers to spoken dialogue in continuous application. Contrary to prior work, where transient spikes in classifier values were found following relevant system actions \cite{Lopes-Dias2019-xb}, the agreement classifier exhibited fluctuations throughout the interaction without clearly time-locked responses to target words. This pattern points to a boundary condition for transferring event-trained classifiers into naturalistic conversation: continuous deployment likely requires mechanisms to identify interaction-relevant moments (e.g., explicit event detection, multimodal cues, or language-tailored feature representations), and it challenges the assumption that evaluative signals are quiescent outside nominal event windows. A further contributor may be construct mismatch. Cursor-based goal-congruency affords discrete toward/away categorizations, whereas linguistic expectedness is distributed and context-dependent across semantic, syntactic, and pragmatic processes \cite{hagoort2000erp}.

The limitations include the small sample size and the exploratory nature of the analyses, and the need to further refine the conversational paradigms. Nevertheless, the present paradigms and pipeline clarify what transfers and where assumptions break down, motivating future work that integrates subjective and behavioral measures to refine workload and agreement decoding in language and ultimately support neural feedback for adaptive conversational AI systems.

\section*{ACKNOWLEDGMENTS}
We thank our colleagues Simon Vogt, Milan Padilla, and Lea Rabe for their valuable feedback and support.


\bibliographystyle{IEEEtran}
\bibliography{bibliography}

@inproceedings{van2016wavenet,
  title={WaveNet: A Generative Model for Raw Audio},
  author={van den Oord, A{\"a}ron and Dieleman, Sander and Zen, Heiga and Simonyan, Karen and Vinyals, Oriol and Graves, Alex and Kalchbrenner, Nal and Senior, Andrew and Kavukcuoglu, Koray},
  booktitle={Proc. SSW 2016},
  pages={125--125},
  year={2016}
}

@article{dinh2025benchmarking,
  title={Benchmarking the Responsiveness of Open-Source Text-to-Speech Systems},
  author={Dinh, Ha Pham Thien and Patamia, Rutherford Agbeshi and Liu, Ming and Cosgun, Akansel},
  journal={Computers},
  volume={14},
  number={10},
  pages={406},
  year={2025},
  publisher={MDPI}
}

@article{casper2023open,
  title={Open problems and fundamental limitations of reinforcement learning from human feedback},
  author={Casper, Stephen and Davies, Xander and Shi, Claudia and Gilbert, Thomas Krendl and Scheurer, J{\'e}r{\'e}my and Rando, Javier and Freedman, Rachel and Korbak, Tomasz and Lindner, David and Freire, Pedro and others},
  journal={arXiv preprint arXiv:2307.15217},
  year={2023}
}

@article{peelle2020completion,
  title   = {Completion norms for 3085 English sentence contexts},
  author  = {Peelle, Jonathan E. and Miller, Ryland L. and Rogers, Chad S. and Spehar, Brent and Sommers, Mitchell S. and Van Engen, Kristin J.},
  journal = {Behavior Research Methods},
  year    = {2020},
  volume  = {52},
  number  = {4},
  pages   = {1795--1799},
  doi     = {10.3758/s13428-020-01351-1}
}

@inproceedings{mikolov2013efficient,
  author    = {Mikolov, Tom{\'{a}}s and Chen, Kai and Corrado, Greg and Dean, Jeffrey},
  title     = {Efficient Estimation of Word Representations in Vector Space},
  booktitle = {1st International Conference on Learning Representations (ICLR 2013), Workshop Track Proceedings},
  year      = {2013},
  url       = {http://arxiv.org/abs/1301.3781}
}

@inproceedings{reimers2019sentencebert,
  title     = {Sentence-bert: Sentence embeddings using siamese bert-networks},
  author    = {Reimers, Nils and Gurevych, Iryna},
  booktitle = {Proceedings of the 2019 Conference on Empirical Methods in Natural Language Processing and the 9th International Joint Conference on Natural Language Processing (EMNLP-IJCNLP)},
  year      = {2019},
  month     = nov,
  address   = {Hong Kong, China},
  publisher = {Association for Computational Linguistics},
  pages     = {3982--3992},
  doi       = {10.18653/v1/D19-1410},
  url       = {https://aclanthology.org/D19-1410/}
}

@article{dimigen2011coregistration,
  title={Coregistration of eye movements and EEG in natural reading: analyses and review.},
  author={Dimigen, Olaf and Sommer, Werner and Hohlfeld, Annette and Jacobs, Arthur M and Kliegl, Reinhold},
  journal={Journal of experimental psychology: General},
  volume={140},
  number={4},
  pages={552},
  year={2011},
  publisher={American Psychological Association}
}

@inproceedings{radford2023robust,
  title={Robust speech recognition via large-scale weak supervision},
  author={Radford, Alec and Kim, Jong Wook and Xu, Tao and Brockman, Greg and McLeavey, Christine and Sutskever, Ilya},
  booktitle={International conference on machine learning},
  pages={28492--28518},
  year={2023},
  organization={PMLR}
}

@article{hamann2022investigating,
  title={Investigating mental workload-induced changes in cortical oxygenation and frontal theta activity during simulated flights},
  author={Hamann, Anneke and Carstengerdes, Nils},
  journal={Scientific Reports},
  volume={12},
  number={1},
  pages={6449},
  year={2022},
  publisher={Nature Publishing Group UK London}
}

@INPROCEEDINGS{Ang2008-sb,
  title     = "Filter bank common spatial pattern ({FBCSP}) in brain-computer
               interface",
  author    = "Ang, Kai Keng and Chin, Zhang Yang and Zhang, Haihong and Guan,
               Cuntai",
  booktitle = "2008 IEEE International Joint Conference on Neural Networks (IEEE
               World Congress on Computational Intelligence)",
  publisher = "IEEE",
  month     =  jun,
  year      =  2008,
  language  = "en"
}

@ARTICLE{Zander2016-rb,
  title     = "Neuroadaptive technology enables implicit cursor control based on
               medial prefrontal cortex activity",
  author    = "Zander, Thorsten O and Krol, Laurens R and Birbaumer, Niels P and
               Gramann, Klaus",
  journal   = "Proc. Natl. Acad. Sci. U. S. A.",
  publisher = "National Academy of Sciences",
  volume    =  113,
  number    =  52,
  pages     = "14898--14903",
  month     =  dec,
  year      =  2016,
  language  = "en"
}

@ARTICLE{Lopes-Dias2019-xb,
  title     = "Online asynchronous decoding of error-related potentials during
               the continuous control of a robot",
  author    = "Lopes-Dias, Catarina and Sburlea, Andreea I and Müller-Putz,
               Gernot R",
  journal   = "Sci. Rep.",
  publisher = "Springer Science and Business Media LLC",
  volume    =  9,
  number    =  1,
  pages     =  17596,
  month     =  nov,
  year      =  2019,
  language  = "en"
}

@ARTICLE{Gherman2025-dl,
  title     = "Towards neuroadaptive chatbots: a feasibility study",
  author    = "Gherman, Diana E and Zander, Thorsten O",
  journal   = "Front. Neuroergonomics",
  publisher = "Frontiers Media SA",
  volume    =  6,
  number    =  1589734,
  pages     =  1589734,
  month     =  oct,
  year      =  2025,
  language  = "en"
}

@ARTICLE{Kothe2024-yu,
  title    = "The Lab Streaming Layer for synchronized multimodal recording",
  author   = "Kothe, Christian and Shirazi, Seyed Yahya and Stenner, Tristan and
              Medine, David and Boulay, Chadwick and Grivich, Matthew I and
              Mullen, Tim and Delorme, Arnaud and Makeig, Scott",
  journal  = "bioRxiv",
  pages    = "2024.02.13.580071",
  month    =  feb,
  year     =  2024,
  language = "en"
}

@ARTICLE{Zander2011-ez,
  title     = "Towards passive brain-computer interfaces: applying
               brain-computer interface technology to human-machine systems in
               general",
  author    = "Zander, Thorsten O and Kothe, Christian",
  journal   = "J. Neural Eng.",
  publisher = "IOP Publishing",
  volume    =  8,
  number    =  2,
  pages     =  025005,
  month     =  apr,
  year      =  2011,
  language  = "en"
}

@ARTICLE{Andreessen2021-oq,
  title     = "Toward neuroadaptive support technologies for improving digital
               reading: a passive {BCI}-based assessment of mental workload
               imposed by text difficulty and presentation speed during reading",
  author    = "Andreessen, Lena M and Gerjets, Peter and Meurers, Detmar and
               Zander, Thorsten O",
  journal   = "User Model. User-adapt Interact.",
  publisher = "Springer Science and Business Media LLC",
  volume    =  31,
  number    =  1,
  pages     = "75--104",
  month     =  mar,
  year      =  2021,
  language  = "en"
}

@INPROCEEDINGS{Krol2016-my,
  title     = "A task-independent workload classifier for neuroadaptive
               technology: Preliminary data",
  author    = "Krol, Laurens R and Freytag, Sarah-Christin and Fleck, Markus and
               Gramann, Klaus and Zander, Thorsten O",
  booktitle = "2016 IEEE International Conference on Systems, Man, and
               Cybernetics (SMC)",
  publisher = "IEEE",
  month     =  oct,
  year      =  2016
}

@TECHREPORT{Palmer2011-og,
  title       = "{AMICA}: An Adaptive Mixture of Independent Component Analyzers
                 with Shared Component",
  author      = "Palmer, Jason A and Kreutz-Delgado, Ken and Makeig, Scott",
  institution = "Swartz Center for Computatonal Neursoscience; University of
                 California San Diego",
  year        =  2011
}

@ARTICLE{Delorme2004-ww,
  title     = "{EEGLAB}: an open source toolbox for analysis of single-trial
               {EEG} dynamics including independent component analysis",
  author    = "Delorme, Arnaud and Makeig, Scott",
  journal   = "J. Neurosci. Methods",
  publisher = "Elsevier BV",
  volume    =  134,
  number    =  1,
  pages     = "9--21",
  month     =  mar,
  year      =  2004,
  language  = "en"
}

@ARTICLE{Kothe2013-et,
  title     = "{BCILAB}: a platform for brain-computer interface development",
  author    = "Kothe, Christian Andreas and Makeig, Scott",
  journal   = "J. Neural Eng.",
  publisher = "IOP Publishing",
  volume    =  10,
  number    =  5,
  pages     =  056014,
  month     =  oct,
  year      =  2013,
  language  = "en"
}

@ARTICLE{Klem1999-td,
  title    = "The ten-twenty electrode system of the International Federation.
              The International Federation of Clinical Neurophysiology",
  author   = "Klem, G H and Lüders, H O and Jasper, H H and Elger, C",
  journal  = "Electroencephalogr. Clin. Neurophysiol. Suppl.",
  volume   =  52,
  pages    = "3--6",
  year     =  1999,
  language = "en"
}

@ARTICLE{Klug2022-bb,
  title    = "The {BeMoBIL} Pipeline for automated analyses of multimodal mobile
              brain and body imaging data",
  author   = "Klug, M and Jeung, S and Wunderlich, A and Gehrke, L and Protzak,
              J and Djebbara, Z and Argubi-Wollesen, A and Wollesen, B and
              Gramann, K",
  journal  = "bioRxiv",
  pages    = "2022.09.29.510051",
  month    =  sep,
  year     =  2022,
  language = "en"
}

@ARTICLE{Klimesch1999-wh,
  title   = "{EEG} alpha and theta oscillations reflect cognitive and memory
             performance: a review and analysis",
  author  = "Klimesch, Wolfgang",
  journal = "Brain Research Reviews",
  volume  =  29,
  number  = "2-3",
  pages   = "169--195",
  month   =  apr,
  year    =  1999
}

@ARTICLE{Blankertz2011-jv,
  title     = "Single-trial analysis and classification of {ERP} components--a
               tutorial",
  author    = "Blankertz, Benjamin and Lemm, Steven and Treder, Matthias and
               Haufe, Stefan and Müller, Klaus-Robert",
  journal   = "Neuroimage",
  publisher = "Elsevier BV",
  volume    =  56,
  number    =  2,
  pages     = "814--825",
  month     =  may,
  year      =  2011,
  language  = "en"
}

@ARTICLE{Pion-Tonachini2019-ea,
  title     = "{ICLabel}: An automated electroencephalographic independent
               component classifier, dataset, and website",
  author    = "Pion-Tonachini, Luca and Kreutz-Delgado, Ken and Makeig, Scott",
  journal   = "Neuroimage",
  publisher = "Elsevier BV",
  volume    =  198,
  pages     = "181--197",
  month     =  sep,
  year      =  2019,
  language  = "en"
}

@ARTICLE{Ziegler2019-rj,
  title         = "Fine-tuning language models from human preferences",
  author        = "Ziegler, Daniel M and Stiennon, Nisan and Wu, Jeffrey and
                   Brown, Tom B and Radford, Alec and Amodei, Dario and
                   Christiano, Paul and Irving, Geoffrey",
  journal       = "arXiv [cs.CL]",
  month         =  sep,
  year          =  2019,
  archivePrefix = "arXiv",
  primaryClass  = "cs.CL"
}

@ARTICLE{Kaufmann2023-ru,
  title         = "A survey of reinforcement learning from human feedback",
  author        = "Kaufmann, Timo and Weng, Paul and Bengs, Viktor and
                   Hüllermeier, Eyke",
  journal       = "arXiv [cs.LG]",
  month         =  dec,
  year          =  2023,
  archivePrefix = "arXiv",
  primaryClass  = "cs.LG",
  language      = "en"
}

@INCOLLECTION{Kaufmann2025-ds,
  title     = "On the challenges and practices of reinforcement learning from
               real human feedback",
  author    = "Kaufmann, Timo and Ball, Sarah and Beck, Jacob and Hüllermeier,
               Eyke and Kreuter, Frauke",
  booktitle = "Communications in Computer and Information Science",
  publisher = "Springer Nature Switzerland",
  address   = "Cham",
  pages     = "276--294",
  series    = "Communications in computer and information science",
  year      =  2025,
  language  = "en"
}

@ARTICLE{Spuler2015-xp,
  title     = "Error-related potentials during continuous feedback: using {EEG}
               to detect errors of different type and severity",
  author    = "Spüler, Martin and Niethammer, Christian",
  journal   = "Front. Hum. Neurosci.",
  publisher = "Frontiers Media SA",
  volume    =  9,
  pages     =  155,
  month     =  mar,
  year      =  2015,
  language  = "en"
}

@ARTICLE{Hassan2025-um,
  title     = "{EEG} workload estimation and classification: a systematic review",
  author    = "Hassan, Jahid and Reza, Shamim and Ahmed, Syed Udoy and Anik,
               Nazmul Haque and Khan, Md Obaydullah",
  journal   = "J. Neural Eng.",
  publisher = "IOP Publishing",
  volume    =  22,
  number    =  5,
  pages     =  051003,
  month     =  oct,
  year      =  2025,
  language  = "en"
}

@ARTICLE{Wickens2008-fl,
  title     = "Multiple resources and mental workload",
  author    = "Wickens, Christopher D",
  journal   = "Hum. Factors",
  publisher = "SAGE Publications",
  volume    =  50,
  number    =  3,
  pages     = "449--455",
  month     =  jun,
  year      =  2008,
  language  = "en"
}

@ARTICLE{Zander2017-rl,
  title     = "Automated task load detection with electroencephalography:
               Towards passive brain–computer interfacing in robotic surgery",
  author    = "Zander, Thorsten O and Shetty, Kunal and Lorenz, Romy and Leff,
               Daniel R and Krol, Laurens R and Darzi, Ara W and Gramann, Klaus
               and Yang, Guang-Zhong",
  journal   = "J. Med. Robot. Res.",
  publisher = "World Scientific Pub Co Pte Ltd",
  volume    =  02,
  number    =  01,
  pages     =  1750003,
  month     =  mar,
  year      =  2017,
  language  = "en"
}

@incollection{billinger2012significant,
  title={Is it significant? Guidelines for reporting BCI performance},
  author={Billinger, Martin and Daly, Ian and Kaiser, Vera and Jin, Jing and Allison, Brendan Z and M{\"u}ller-Putz, Gernot R and Brunner, Clemens},
  booktitle={Towards Practical Brain-Computer Interfaces: Bridging the Gap from Research to Real-World Applications},
  pages={333--354},
  year={2012},
  publisher={Springer}
}

@article{ling2021factors,
  title={Factors influencing users' adoption and use of conversational agents: A systematic review},
  author={Ling, Erin Chao and Tussyadiah, Iis and Tuomi, Aarni and Stienmetz, Jason and Ioannou, Athina},
  journal={Psychology \& marketing},
  volume={38},
  number={7},
  pages={1031--1051},
  year={2021},
  publisher={Wiley Online Library}
}

@article{hagoort2000erp,
  title={ERP effects of listening to speech: semantic ERP effects},
  author={Hagoort, Peter and Brown, Colin M},
  journal={Neuropsychologia},
  volume={38},
  number={11},
  pages={1518--1530},
  year={2000},
  publisher={Elsevier}
}

\end{document}